\documentclass{article}
\usepackage{amsfonts}
\usepackage{amssymb}
\usepackage{amsmath}

\newtheorem{theorem}{Theorem}

\newtheorem{definition}[theorem]{Definition}

\newtheorem{remark}[theorem]{Remark}

\newenvironment{acknowledgement}
{\par\noindent\textbf{Acknowledgement.}\quad}
{\par}

\input{tcilatex}

\begin{document}

\author{Naichung Conan Leung and Mengqi Yang}
\title{Mirror symmetry in 3d \\
in 3d mirror symmetry\\
}
\maketitle

\begin{abstract}
Given a compact CY3 $Y$, its A-side (resp. B-side) universal intermediate
Jacobian $X$ (resp. $X^{!}$) admits a natural hyperkahler structure. Both $X$
and $X^{!}$ determines 3d Rozansky-Witten theories, in both A-model and
B-model. We describe some surprising 3d mirror symmetry phenomena between $X$
and $X^{!}$.

This includes (i) a 3d SYZ construction of $X^{!}$ from $X$ via moduli of
certain 3d A-branes in $X$ constructed from $D^{b}\left( Y,\Omega \right) $
with varying stability conditions given by $\varpi $; (ii) The 3d B-brane on 
$X^{!}$, constructed from varying $D^{b}\left( Y,\Omega \right) $ with fixed 
$\varpi $, is realized as the 3d SYZ transformation of a cotangent fiber 3d
A-brane in $X$.

Under mirror symmetry between $Y$ and $Y^{\vee }$, the roles for $X$ and $%
X^{!}$ got interchanged. 

We also construct 3d A- and B-branes in $X$ and $X^{!}$ via DT theory and
discrete symmetries on $Y$ and $Y^{\vee }$.
\end{abstract}

Given a compact Calabi-Yau manifold $Y$ of complex dimension three (abbrev.
CY3) with K\"{a}hler form $\omega $ and holomorphic volume form $\Omega 
\footnote{%
Complex structure $J$ on $Y$ is determined by $\Omega $.}$, the moduli space 
$\mathcal{M}_{cpx}\left( Y\right) $ of $\Omega $ admits a special K\"{a}hler
structure \cite{Freed}. This implies that the universal intermediate
Jacobian $X^{!}=T_{\mathbb{Z}}^{\ast }\mathcal{M}_{cpx}\left( Y\right)
\backslash T^{\ast }\mathcal{M}_{cpx}\left( Y\right) $ is hyperk\"{a}hler.
The moduli space\footnote{%
	Enlarged to include all A-models on $Y$.}  $\mathcal{M}_{sympl}\left( Y\right) $ of complexified K\"{a}hler structures $ \varpi $ on $Y$ with quantum corrections is
also special K\"{a}hler, and the corresponding A-sided universal
intermediate Jacobian $X=T_{\mathbb{Z}}^{\ast }\mathcal{M}_{sympl}\left(
Y\right) \backslash T^{\ast }\mathcal{M}_{sympl}\left( Y\right) $ is also
hyperk\"{a}hler.

Closed string mirror symmetry conjecture predicts that between mirror CYn
manifold $Y$ and $Y^{\vee }$, we have%
\begin{equation*}
\mathcal{M}_{sympl}\left( Y\right) \simeq \mathcal{M}_{cpx}\left( Y^{\vee
}\right) \text{ ~and }\mathcal{M}_{cpx}\left( Y\right) \simeq \mathcal{M}%
_{sympl}\left( Y^{\vee }\right) ,
\end{equation*}%
as Frobenius manifolds. When $n=3,$ we should have an isomorphism between $X$
for $Y$ and $X^{!}$ for $Y^{\vee }$ as hyperk\"{a}hler manifolds.

 Open string mirror symmetry conjecture, or the
Kontsevich HMS conjecture, predicts that under the mirror map $\varpi \longleftrightarrow \Omega ^{\vee }$ and $\Omega
\longleftrightarrow \varpi ^{\vee }$, there is a quasi-equivalence between derived
category of coherent sheaves on $\left( Y,\Omega \right) $ with stability
condition determined by $\varpi $ and Fukaya category of Lagrangians on $%
\left( Y^{\vee },\varpi ^{\vee }\right) $ with stability condition
determined by $\Omega ^{\vee }$. We denote this as%
\begin{equation*}
\left( D^{b}\left( Y,\Omega \right) ,\varpi \right) \simeq \left( Fuk\left(
Y^{\vee },\varpi ^{\vee }\right) ,\Omega ^{\vee }\right) .\text{ }
\end{equation*}%
Similarly, we have%
\begin{equation*}
\left( Fuk\left( Y,\varpi \right) ,\Omega \right) \simeq \left( D^{b}\left(
Y^{\vee },\Omega ^{\vee }\right) ,\varpi ^{\vee }\right) .\text{ }
\end{equation*}
\qquad

Strominger-Yau-Zaslow proposed a geometric explanation of this mysterious duality in \cite{SYZ}. This SYZ conjecture says that (i) \textit{SYZ construction}: $Y^{\vee }$ is the
moduli space of certain A-branes in $Y$ and (ii) \textit{SYZ\ transformation}%
: the mirror functor in HMS is given by a fiberwise Fourier transformation.

\bigskip

In physics, hyperk\"{a}hler manifolds $X$ and $X^{!}$ define 3d RW-theory,
in both A-model and B-model. Fixing $\Omega $ on $Y$, $D^{b}\left( Y,\Omega
\right) $ with varying $\varpi \in \mathcal{M}_{sympl}\left( Y\right) $
defines a 3d A-brane $\underline{_{A}\mathcal{C}_{\Omega }}$ in $X$. If all
deformation of $\underline{_{A}\mathcal{C}_{\Omega }}$ is of the same form,
then $X^{!}$ can be realized as given by a 3d SYZ construction on $X$. On
the other hand, if we fix $\varpi $ and view it as a stability condition on $%
D^{b}\left( Y,\Omega \right) $ with varying $\Omega \in \mathcal{M}%
_{cpx}\left( Y\right) $, then this defines a 3d B-brane $\underline{_{B}%
\mathcal{D}_{\varpi }}$ on $X^{!}$. We will argue that $\underline{_{B}%
\mathcal{D}_{\varpi }}$ is the 3d SYZ transformation of a cotangent fiber 3d
A-brane on $X$. Even though we are not claiming that $X$ and $X^{!}$ and 3d
mirror to each other, we have just demonstrate some surprising 3d mirror
behaviors between $X$ and $X^{!}$.

If we use $Fuk\left( Y,\varpi \right) $, instead of $D^{b}\left( Y,\Omega
\right) $, or equivalently we consider the mirror CY3 $Y^{\vee }$ instead of 
$Y$, then the roles of $X$ and $X^{!}$ got interchanged.

In the last two sections, we will construct new 3d A- and B-branes on $X$
and $X^{!}$ using Abelian symmetries of $Y$ and categorical DT theory.
Readers should be warned that most of the discussion in this article are
conjectural in nature from a mathematical point of view. Thus we have assume
various mirror symmetry conjectures holds, which include closed string mirror symmetry
conjecture, open string mirror symmetry conjecture, existence of categorical DT-invariants
in both A- and B-models and their equivalence under mirror symmetry and so
on.

\bigskip

\begin{acknowledgement}
This research is substantially supported by grants from the Research Grants
Council of the Hong Kong Special Administrative Region, China (Project No.
CUHK14306322, CUHK14305923 and CUHK14302224) and direct grants from the
Chinese University of Hong Kong. We thank Pierrick Bousseau,
Kifung Chan, Will Donovan, Ziming Ma, Michael McBreen, Mauricio Romo  and Shing-Tung Yau for useful discussions.
\end{acknowledgement}

\bigskip

\section{Mirror symmetry}

Given a compact Calabi-Yau manifold $Y$ of complex dimension $n$, abbrev.
CYn, with $ t\in \mathbb{C}^\times $ scaled complexified K\"{a}hler form $\varpi=te^{\omega +iB} $ and holomorphic volume form $\Omega $.
For simplicity, we assume that $Y$ is a strict CY, i.e. its holonomy group
equals $SU\left( n\right) $, in particular, $H^{n,0}\left( Y\right) \simeq 
\mathbb{C}$ and $H^{k,0}\left( Y\right) =0$ for any $k$ between $0$ and $n$.
In physics, there are two topological twists of the superstring theory with
target $Y$, called the A-model and B-model. They correspond to the
symplectic geometry and the complex geometry of $Y$ respectively in mathematics.

Mirror symmetry conjecture predicts that there is another CYn $\left(
Y^{\vee },\varpi ^{\vee },\Omega ^{\vee }\right) $ where the A-model (resp.
B-model) of $Y$ is equivalent to the B-model (resp. A-model) of $Y^{\vee }$.

B-models on $Y$ are parametrized by its complex structure $J$ together with
a choice of a $J$-holomorphic volume form $\Omega $, which is unique up to $%
\mathbb{C}^{\times }$-scaling. Locally $\Omega =dz_{1}\wedge \cdots \wedge
dz_{n}$ on a coordinate chart $U\subset \mathbb{C}^{n}$. In fact, every
closed decomposable forms $\Omega \in \Omega ^{n}\left( Y,\mathbb{C}\right) $
with nonvanishing $\Omega \wedge \overline{\Omega }$ determines a complex
structure $J$ on $Y$ in a way that $\phi \in \left( T^{\ast }Y\right) ^{1,0}$
is characterized by $\phi \wedge \Omega =0$. Furthermore $\Omega $ is a $J$%
-holomorphic volume form in $Y$. Thus the Teichmuller space and the moduli
space of B-models on $Y$ is given by 
\begin{eqnarray*}
\mathcal{T}_{cpx}\left( Y\right)  &=&\left\{ \Omega \in \Omega ^{n}\left( Y,%
\mathbb{C}\right) |\text{decomposable, }\Omega \wedge \overline{\Omega }%
\text{ nonvanishing, }d\Omega =0\right\} /Diff^{0}\left( Y\right) \text{,} \\
\mathcal{M}_{cpx}\left( Y\right)  &=&\left\{ \Omega \in \Omega ^{n}\left( Y,%
\mathbb{C}\right) |\text{decomposable, }\Omega \wedge \overline{\Omega }%
\text{ nonvanishing, }d\Omega =0\right\} /Diff\left( Y\right) \text{,}
\end{eqnarray*}%
respectively, where $Diff^{0}\left( Y\right) $ is the identity component of
the group of diffeomorphisms $Diff\left( Y\right) $ of $Y$.

First order deformations of $J$ are parametrized by $H^{1}\left( Y,TY\right) 
$, which is isomorphic to $H^{1}\left( Y,\wedge ^{n-1}T^{\ast }Y\right)
\simeq H^{n-1,1}\left( Y\right) $ via contracting with $\Omega $. Every
first order deformation of $J$ is unobstructed \cite{Tian}\cite{Todorov}.
Thus $\mathcal{T}_{cpx}\left( Y\right) $ and $\mathcal{M}_{cpx}\left(
Y\right) $ are smooth complex manifolds of dimension $1+h^{n-1,1}\left(
Y\right) $. The action of $Diff^{0}\left( Y\right) $ on $H^{\ast }\left( Y,%
\mathbb{C}\right) $ is trivial, thus we have the period map%
\begin{eqnarray*}
\mathcal{P} &:&\mathcal{T}_{cpx}\left( Y\right) \rightarrow H^{n}\left( Y,%
\mathbb{C}\right) \\
\mathcal{P}\left( \Omega \right) &=&\left[ \Omega \right] \text{,}
\end{eqnarray*}%
whose differential is always injective by the local Torelli theorem. Namely $%
\mathcal{T}_{cpx}\left( Y\right) $ is an immersed complex submanifold of $%
H^{n}\left( Y,\mathbb{C}\right) $ whose tangent space at $\Omega $ is given
by $H^{n,0}\left( Y\right) \oplus H^{n-1,1}\left( Y\right) \subset
H^{n}\left( Y,\mathbb{C}\right) $ in the Hodge decomposition for the complex
structure $J$ determined by $\Omega $. The trivial bundle $\underline{%
H^{n}\left( Y,\mathbb{C}\right) }$ over $\mathcal{T}_{cpx}\left( Y\right) $
descends to a flat bundle over $\mathcal{M}_{cpx}\left( Y\right) $ with
Gauss-Manin connection.

\bigskip

On the other hand, the moduli space $\mathcal{M}_{sympl}\left( Y\right) $ of
A-models does not have a similar global description in general, with the exception when $ Y $ has a gauge linear sigma-model (abbrev. GLSM) description. Near a large volume
limit (abbrev. LVL) point, $\mathcal{M}_{sympl}\left( Y\right) $ would
admits a local chart parametrized by complexified K\"{a}hler forms $\varpi$ on $ Y $, which receives quantum corrections given by genus zero
three points Gromov-Witten invariants on $Y$. However, different LVLs may arise from different CY manifolds. For example, if $Y^{\prime }$ is another CY3, birational to $Y$, then its complexified K\"{a}hler forms will give a coordinate chart around another LVL point in $\mathcal{M}_{sympl}\left( Y\right) $. Complex dimension of $\mathcal{M}_{sympl}\left( Y\right) $
equals $1+h^{1,1}\left( Y\right) $.

When $ Y $ admits a GLSM description, $\mathcal{M}_{sympl}\left( Y\right) $ can be identified globally. Concretely, suppose $ \mathbb{C}^{N} $ admits a linear Hamiltonian $ T^k $-action with moment map $ \mu: \mathbb{C}^N \to \mathbb{R}^k $ and $ W:\mathbb{C}^N \to \mathbb{C} $ is a $ T^k $-invariant polynomial such that
\begin{eqnarray*}
	Y=\mathrm{Crit}(W)/\!\!/_{\lambda} T^k
\end{eqnarray*}
for any $ \lambda\in\mathbb{R}^k $. Then
\begin{eqnarray*}
	\mathcal{M}_{sympl}\left( Y\right) \cong \mathbb{C}^{\times k}\setminus\Delta
\end{eqnarray*}
for some explicit complex hypersurface $ \Delta\subset \mathbb{C}^{\times k} $.
For example, when $ S^1 $ acting on $ \mathbb{C}^6 $ with weights $ (-5,1,1,1,1,1) $ and
\begin{eqnarray*}
	W(p,z_0,...,z_4)=p f_5(z_0,...,z_4)
\end{eqnarray*}
for a degree five homogeneous polynomial $ f_5 $, then $ \mathrm{Crit}(W)/\!\!/_{1} S^1 $ is a quintic CY3 inside $ \mathbb{CP}^{4} $ and
\begin{eqnarray*}
	\mathcal{M}_{sympl}\left( Y\right) =\mathbb{C}^\times\setminus\Delta
\end{eqnarray*}
with $ \Delta=\{(-5)^5\} $.

\begin{remark}
For semiflat mirror CYn pair $Y$ and $Y^{\vee }$, explicit SYZ
transformations interchange $\varpi $ with $\Omega ^{\vee }$ and $\Omega $
with $\varpi ^{\vee }$ \cite{LYZ}.
\end{remark}

By abuse of notations, we will simply denote an element in $\mathcal{M}%
_{sympl}\left( Y\right) $ as $\varpi $ and called it a symplectic structure
on $Y$ (or a $\Pi $-stability condition of $D^{b}\left( Y,\Omega \right) $
as will be explained later).

\bigskip

Closed string mirror symmetry conjecture predicts that 
\begin{eqnarray*}
&&\mathcal{M}_{cpx}\left( Y\right) \overset{\simeq }{\longrightarrow }%
\mathcal{M}_{sympl}\left( Y^{\vee }\right) , \\
&&\mathcal{M}_{sympl}\left( Y\right) \overset{\simeq }{\longrightarrow }%
\mathcal{M}_{cpx}\left( Y^{\vee }\right) .
\end{eqnarray*}%
There maps are called mirror maps and inverse mirror maps respectively. The
mirror map relates periods of $Y$ with genus zero three points Gromov-Witten
invariants on $Y^{\vee }$.

\bigskip

For open strings, supersymmetric (SUSY) boundary conditions are called
branes, which are certain submanifolds of $Y$ coupled with a family of 1d
TFTs, or equivalently a family of Hilbert spaces. An A-brane\footnote{%
There are more general coisotropic A-branes defined by Kapustin and Orlov
[KO].} is a Lagrangian submanifold $L$ in $Y$ coupled with a flat family of
Hilbert spaces\footnote{%
Hilbert spaces are regarded as SUSY quantum mechanics, or 1d TFTs.}, namely
a unitary flat bundle $E$ over $L$. A B-brane\footnote{%
More generally a complex of coherent sheaves on $Y$.} is a complex
submanifold $L$ in $Y$ coupled with a holomorphic family of Hilbert spaces,
namely a Hermitian holomorphic vector bundle $E$ over $L$. We denote such a
pair as $\underline{E}=\left( L,E\right) $.

Morphisms, or Hom, between branes can be interpreted as the intersection
theory between $\underline{E_{1}}$ and $\underline{E_{2}}$, which are 
\begin{equation*}
_{A}Hom\left( \underline{E_{1}},\underline{E_{2}}\right) =HF\left( 
\underline{E_{1}},\underline{E_{2}}\right) ,
\end{equation*}%
the Floer homology of Lagrangian branes in Floer theory for A-model of the
symplectic manifold $Y$ and%
\begin{equation*}
_{B}Hom\left( \underline{E_{1}},\underline{E_{2}}\right) =Ext_{O_{Y}}^{\ast
}\left( \underline{E_{1}},\underline{E_{2}}\right) ,
\end{equation*}%
the extension group in algebraic geometry for B-model of the complex
manifold $Y$.

These structures define the Fukaya category $Fuk\left( Y,\varpi \right) $ in
A-model and derived category of coherent sheaves $D^{b}\left( Y,\Omega
\right) $ in B-model.

\bigskip

Open string mirror symmetry conjecture says that under the mirror symmetry
between $\left( Y,\varpi ,\Omega \right) $ and $\left( Y^{\vee },\varpi
^{\vee },\Omega ^{\vee }\right) $, we have a quasi-equivalence of triangular
categories%
\begin{equation*}
\mathcal{F}:Fuk\left( Y,\varpi \right) \overset{\simeq }{\longrightarrow }%
D^{b}\left( Y^{\vee },\Omega \right) .
\end{equation*}%
This is Kontsevich's Homological Mirror Symmetry (HMS) conjecture \cite%
{Kontsevich ICM}.

Strominger-Yau-Zaslow \cite{SYZ} proposed that (i) the mirror manifold $%
Y^{\vee }$ to $Y$ is the moduli space of Lagrangian A-branes $\underline{E}%
=\left( L,E\right) $ in $Y$ with $L$ a torus and $rank\left( E\right) =1$;
(ii) The above mirror functor $\mathcal{F}$ should be given by T-duality, or
mathematically speaking family Floer theory \cite{Abouzaid}\cite{Cho Hong
Lau}. Very loosely speaking, the stalk of $\mathcal{F}\left( \underline{E}%
^{\prime }\right) $ over $\underline{E}\in Y^{\vee }$ is given by $HF\left( 
\underline{E}^{\prime },\underline{E}\right) $. We call this the SYZ
transformation.

\begin{remark}
When $Y$ is a Fano manifold, there is a holomorphic function, called the
potential function, 
\begin{equation*}
W^{\vee }:Y^{\vee }=\left\{ \underline{E}=\left( L,E\right) \text{ in }%
Y\right\} \rightarrow \mathbb{C}
\end{equation*}%
obtained by counting Maslov two holomorphic disks in $Y$ with boundary lying
on $L$. When $Y$ is Calabi-Yau, then $W^{\vee }=0$. Superstring theory for $%
\left( Y^{\vee },W^{\vee }\right) $ is called a LG model. Mathematically
speaking, $\left( Y^{\vee },W^{\vee }\right) $ determines a $\left(
-1\right) $-shifted complex symplectic manifold \cite{PPTV}, which is $%
T^{\ast }\left[ -1\right] Y^{\vee }$ when $W^{\vee }=0$. Furthermore,
B-brane $\underline{E^{\vee }}=\left( L^{\vee },E^{\vee }\right) $ in $%
Y^{\vee }$ determines $\left( -1\right) $-shifted complex Lagrangian in $%
T^{\ast }\left[ -1\right] Y^{\vee }$ with supported on $N^{\ast }\left[ -1%
\right] _{L^{\vee }/Y^{\vee }}$. Thus mirror symmetry could be interpreted
as a duality between $0$-shifted real symplectic geometry of $Y$ and $\left(
-1\right) $-shifted complex symplectic geometry of $T^{\ast }\left[ -1\right]
Y^{\vee }$, with $Y^{\vee }=\left\{ \underline{E}=\left( L,E\right) \text{
in }Y\right\} $. Such a SYZ perspective will be crucial in the construction
of 3d mirror later.
\end{remark}

\bigskip

In the B-model on $\left( Y,\varpi ,\Omega \right) $ in physics, $\varpi $
plays the role of $\Pi $-stability by the work of Douglas \cite{Douglas}.
This notion was generalized mathematically by Bridgeland \cite{Bridgeland},
called the stability condition of the triangulated category $D^{b}\left(
Y,\Omega \right) $ of boundary conditions of the B-model. The space $%
Stab\left( D^{b}\left( Y,\Omega \right) \right) $ of stability conditions of 
$D^{b}\left( Y,\Omega \right) $ is a complex manifold, locally modelled on $%
H^{ev}\left( Y,\mathbb{C}\right) $. It is expected that $\mathcal{M}%
_{sympl}\left( Y\right) $ is a closed complex submanifold of $Stab\left(
D^{b}\left( Y,\Omega \right) \right) $, locally modelled on $\mathbb{C}%
\oplus H^{2}\left( Y,\mathbb{C}\right) $. Similarly, in the A-model on $%
\left( Y,\varpi ,\Omega \right) $, $\Omega $ plays the role of $\Pi $%
-stability in physics and we expect that $\mathcal{M}_{cpx}\left( Y\right)
\subset Stab\left( Fuk\left( Y,\varpi \right) \right) $. Therefore, we also
call an element in $\mathcal{M}_{cpx}\left( Y\right) $ (resp. $\mathcal{M}%
_{sympl}\left( Y\right) $) a $\Pi $-stability condition on $Fuk\left(
Y,\varpi \right) $ (resp. $D^{b}\left( Y,\Omega \right) $). \bigskip 

\section{Closed string mirror symmetry in 3d}

When $Y$ is a Calabi-Yau manifold of dimension three, i.e. a CY3, the wedge
product 
\begin{eqnarray*}
H^{3}\left( Y,\mathbb{C}\right) \otimes H^{3}\left( Y,\mathbb{C}\right) 
&\longrightarrow &\mathbb{C} \\
\left( \phi ,\eta \right)  &\longmapsto &\int_{Y}\phi \wedge \eta \text{,}
\end{eqnarray*}%
defines a linear complex symplectic structure on $H^{3}\left( Y,\mathbb{C}%
\right) $ and via the period map $\mathcal{M}_{cpx}\left( Y\right) $ becomes
a complex Lagrangian submanifold in $H^{3}\left( Y,\mathbb{C}%
\right) $ locally, as its tangent space at $\Omega \in \mathcal{M}_{cpx}\left(
Y\right) $ is $H^{3,0}\left( Y\right) \oplus H^{2,1}\left( Y\right) $ inside%
\begin{eqnarray*}
H^{3}\left( Y,\mathbb{C}\right)  &=&H^{3,0}\left( Y\right) \oplus
H^{2,1}\left( Y\right) \oplus H^{1,2}\left( Y\right) \oplus H^{0,3,0}\left(
Y\right)  \\
&=&H^{3,0}\left( Y\right) \oplus H^{2,1}\left( Y\right) \oplus \overline{%
H^{3,0}\left( Y\right) \oplus H^{2,1}\left( Y\right) } \\
&=&T^{\ast }\left( H^{3,0}\left( Y\right) \oplus H^{2,1}\left( Y\right)
\right) \ 
\end{eqnarray*}%
by the Hodge decomposition for compact K\"{a}hler manifolds.

The intermediate Jacobian of a projective threefold $\left( Y,\Omega \right) 
$ is defined as%
\begin{equation*}
\mathcal{J}_{cpx}\left( Y,\Omega \right) =H^{3}\left( Y,\mathbb{Z}\right)
\backslash H^{3}\left( Y,\mathbb{C}\right) /H^{3,0}\left( Y\right) \oplus
H^{2,1}\left( Y\right) .
\end{equation*}%
Similar to the usual Jacobian $Jac\left( Y,\Omega \right) =H^{1}\left( Y,%
\mathbb{Z}\right) \backslash H^{1}\left( Y,\mathbb{C}\right) /H^{1,0}\left(
Y\right) $, $\mathcal{J}_{cpx}\left( Y,\Omega \right) $ is a principally
polarized Abelian variety. By putting all $\mathcal{J}_{cpx}\left( Y,\Omega
\right) $'s together, we have the universal intermediate Jacobian $\mathcal{J%
}_{cpx}\left( Y\right) $ together with a fibration 
\begin{equation*}
\pi _{cpx}:\mathcal{J}_{cpx}\left( Y\right) \rightarrow \mathcal{M}%
_{cpx}\left( Y\right) \text{,}
\end{equation*}%
with fiber $\pi _{cpx}^{-1}\left( \Omega \right) \simeq \mathcal{J}%
_{cpx}\left( Y,\Omega \right) $. From the above discussion, we have 
\begin{equation*}
\mathcal{J}_{cpx}\left( Y\right) =T_{\mathbb{Z}}^{\ast }\mathcal{M}%
_{cpx}\left( Y\right) \backslash T^{\ast }\mathcal{M}_{cpx}\left( Y\right)
\end{equation*}%
with a lattice subbundle $T_{\mathbb{Z}}^{\ast }\mathcal{M}_{cpx}\left(
Y\right) \subset T^{\ast }\mathcal{M}_{cpx}\left( Y\right) $ defined by $%
H^{3}\left( Y,\mathbb{Z}\right) \subset H^{3}\left( Y,\mathbb{C}\right) $.
In particular, $\mathcal{J}_{cpx}\left( Y\right) $ is a holomorphic
symplectic manifold and $\pi _{cpx}$ is a complex Lagrangian fibration with
section.

As a matter of fact, $\mathcal{J}_{cpx}\left( Y\right) $ admits a natural
hyperk\"{a}hler structure, which arises from the special K\"{a}hler
structure on $\mathcal{M}_{cpx}\left( Y\right) $ for CY3 \cite{Freed}.

\bigskip 

There is a similar special K\"{a}hler structure on the complexified K\"{a}%
hler moduli space\footnote{%
With quantum corrections given by genus zero GW invariants.} and hyperk\"{a}%
hler structure on the A-sided universal intermediate Jacobian. It is
expected that these structures extend to the whole moduli space of A-models,
i.e. 
\begin{equation*}
\pi _{sympl}:\mathcal{J}_{sympl}\left( Y\right) \rightarrow \mathcal{M}%
_{sympl}\left( Y\right) \text{,}
\end{equation*}%
a complex Lagrangian fibration with section on a hyperk\"{a}hler manifold $%
\mathcal{J}_{sympl}\left( Y\right) =T_{\mathbb{Z}}^{\ast }\mathcal{M}%
_{sympl}\left( Y\right) \backslash T^{\ast }\mathcal{M}_{sympl}\left(
Y\right) $.

Closed string mirror symmetry for CYn predicts that $\mathcal{M}%
_{sympl}\left( Y\right) \overset{\simeq }{\longrightarrow }\mathcal{M}%
_{cpx}\left( Y^{\vee }\right) .$ When $n=3$, this should lift to an
isomorphism 
\begin{equation*}
\mathcal{J}_{sympl}\left( Y\right) \overset{\simeq }{\longrightarrow }%
\mathcal{J}_{cpx}\left( Y^{\vee }\right) \text{,}
\end{equation*}%
as hyperk\"{a}hler manifolds with compatible complex Lagrangian fibration
structures,

\begin{equation*}
\begin{array}{ccc}
\mathcal{J}_{sympl}\left( Y\right)  & \overset{\simeq }{\longrightarrow } & 
\mathcal{J}_{cpx}\left( Y^{\vee }\right)  \\ 
\text{ }\downarrow ~\pi _{sympl} &  & \text{ }\downarrow \pi _{cpx}^{\vee }
\\ 
\mathcal{M}_{sympl}\left( Y\right)  & \overset{\simeq }{\longrightarrow } & 
\mathcal{M}_{cpx}\left( Y^{\vee }\right) 
\end{array}.%
\end{equation*}%
We would also have the same duality with the roles of $Y$ and $Y^{\vee }$ got exchanged. 

\bigskip

Remark: In either A- or B-model, $\mathcal{M}\left( Y\right) $ and $\mathcal{%
J}\left( Y\right) $ are Coulomb branches of a 10 dimensional SUSY theory
compactified to 4 dimension and 3 dimension respectively. The low energy
effective theory would then becomes a 3d $\sigma $-model with target $%
\mathcal{J}\left( Y\right) $.

\bigskip 

Open string mirror symmetry for CYn relates Lagrangian submanifolds in $Y$
to coherent sheaves on $Y^{\vee }$. When $n=3$, there are (categorical) DT
counts for these geometric objects and such dualities will be discussed in
later sections. 

\section{3d mirror symmetry}

The physical theory of 3d N=4 SUSY $\sigma $-model on a hyperk\"{a}hler
manifold $X$ was constructed by Rozansky and Witten \cite{Rozansky Witten} as a B-model,
called the RW B-model. 3d N=4 super Yang-Mills theory associated to $%
\left( G,Y\right) $, with $G$ a compact Lie group $G$ and $V$ a complex
representation of $G$, its low energy effective field theory can be viewed as a RW theory on the hyperk\"{a}hler
stack $T^{\ast }\left[ V/G_{\mathbb{C}}\right] $. 

Recall that a Riemannian manifold $\left( X,g\right) $ of real dimension $4n$
is hyperk\"{a}hler if the holonomy group of its Levi-Civita connection is a
subgroup of $Sp\left( n\right) $. It admits three K\"{a}hler structures $%
\omega _{I}$, $\omega _{J}$ and $\omega _{K}$ satisfying the Hamilton
relation $I^{2}=J^{2}=K^{2}=IJK=-id$. Fixing any one such complex structure,
say $J$, then%
\begin{equation*}
\omega ^{\mathbb{C}}=\omega _{I}+i\omega _{K}
\end{equation*}%
defines a $J$-holomorphic symplectic structure on $X$. Kapranov \cite%
{Kapranov} showed that the B-model RW theory depends only on such a
holomorphic symplectic structure on $X$.

Kapustin and Vyes \cite{Kapustin Vyas} proposed a physical A-model RW theory
on any 1-shifted real symplectic manifold $X_{\mathbb{R}}$ \cite{PPTV}. A
construction of the corresponding A-model 2-category was proposed in \cite%
{Pascaleff}. On the other hand, if $X$ is a hyperk\"{a}hler manifold with $%
Sp\left( 1\right) $-symmetry, physicists also constructed a 3d A-model for $X
$, which is viewed as a 3d generalized Seiberg-Witten theory. When $%
X=T^{\ast }C$, then the 1-shifted cotangent bundle $X_{\mathbb{R}}=T^{\ast }%
\left[ 1\right] C$ is a 1-shifted real symplectic manifold $X_{\mathbb{R}}$
and we suspect that these two 3d A-models are closely related to each other.

3d mirror symmetry was first proposed by Intriligator and N. Seiberg for 3d
N=4 super Yang-Mills theory \cite{Intriligator Seiberg}. Symplectic duality
among hyperk\"{a}hler resolutions was later discovered by Braden, Licata,
Proudfoot and Webster \cite{BLPW} and realized as a 3d mirror symmetry
phenomena. In the gauge theory setting, a general construction of 3d mirror
to $T^{\ast }\left[ V/G_{\mathbb{C}}\right] $ was proposed by Braverman,
Finkelberg and Nakajima \cite{BFN} and many 3d mirror phenomena were verfied
for such 3d mirror pairs. 3d homological mirror symmetry predicts that the
2-categories of dual 3d mirror manifolds are equivalent to each other by
exchanging A-side and B-side. This is verified by Gammage and Hilburn in the
hypertoric cases \cite{Gammage Hilburn}. A SYZ type proposal to explain 3d
mirror symmetry and relate it to 2d mirror symmetry was given Chan and the
first author \cite{Chan Leung 3d}\cite{Chan Leung}.

\bigskip 

A physical boundary condition, or a physical 3d brane, for the 3d RW theory on $\left( X,\omega ^{%
\mathbb{C}}\right) $, in either A-model or B-model, is a $\omega ^{\mathbb{C}%
}$-complex Lagrangian $C$ in $X$ coupled with a \textit{certain} family $%
\left\{ \left( \mathcal{Y}_{c},\sigma _{c}\right) \right\} _{c\in C}$ of 2d
TFT over $C$. $ C $ is called the support of the brane. The datum we need for a 2d TFT consists of a category $%
\mathcal{Y}_{c}$ of its 2d boundary conditions together with a $\Pi $%
-stability condition $\sigma _{c}$ of $\mathcal{Y}_{c}$.

Given a K\"{a}hler manifold $\left( Y,\varpi \right) $ with a CY-fibration 
\begin{equation*}
\pi :Y\rightarrow C
\end{equation*}%
with a relative holomorphic volume form $\Omega _{Y/C}\in H^{0}\left(
Y,\omega _{Y/C}\right) $, where $\omega _{Y/C}$ is the relative dualizing
sheaf, and $C\subset X$ is a $\omega ^{\mathbb{C}}$-complex Lagrangian
submanifold in $X$. This gives a family of CY varieties$\ \left\{ \left(
Y_{c},\varpi _{c},\Omega _{c}\right) \right\} _{c\in C}$ over $C$. Over the
locus $C_{sm}\subset C$ of smooth fibers, $\left\{ Y_{c}\right\} _{c\in
C_{sm}}$ is a locally trivial family of smooth manifolds, then their
cohomology groups $H^{\ast }\left( Y_{c},\mathbb{C}\right) $'s form a flat
bundle $R^{\bullet }\pi _{\ast }\underline{\mathbb{C}}$ over $C_{sm}$.
Furthermore $\left\{ \left[ \varpi _{c}\right] \right\} _{c\in C}$ is a flat
section of $R^{\bullet }\pi _{\ast }\underline{\mathbb{C}}$ because of $%
d\varpi =0$.

This implies that $\left\{ Fuk\left( Y_{c},\varpi _{c}\right) \right\}
_{c\in C}$ is a flat family of categories over $C_{sm}$. We expect that it
has mild singularities along $C\backslash C_{sm}$ and give a perverse
schober \cite{Kapranov} over $C$. As $\pi $ is a CY-fibration, the $\Pi $%
-stability conditions $\Omega _{c}$ on $Fuk\left( Y_{c},\varpi _{c}\right) $
varies holomorphically over $C$.

For B-model considerations, the family of 2d TFTs $\left\{ D^{b}\left(
Y_{c},\Omega _{c}\right) \right\} _{c\in C}$ is a holomorphic family of
triangulated dg-categories together with a flat family of $\Pi $-stability conditions $%
\left\{ \varpi _{c}\right\} _{c\in C}$. Note that a holomorphic family of
categories over $C$ is the same as a $D^{b}\left( C\right) $-linear
category, which is $D^{b}\left( Y\right) $ in this example. The above
discussion motivates the following definition of a physical 3d brane.

\bigskip 

\begin{definition}
A physical 3d A-brane (resp. B-brane) in a holomorphic symplectic manifold $%
\left( X,\omega ^{\mathbb{C}}\right) $ is a flat (resp. holomorphic) family
of triangulated dg-categories $\mathcal{Y}_{c}$'s equipped with a holomorphic
(resp. flat) family of $\Pi $-stability conditions $\sigma _{c}$ of $%
\mathcal{Y}_{c}$ over a $\omega ^{\mathbb{C}}$-complex Lagrangian
submanifold $C$ in $X$. We will denote it as $\left\{ \left( \mathcal{Y}%
_{c},\sigma _{c}\right) \right\} _{c\in C}$; $\pi _{\mathcal{Y}}:\left( 
\mathcal{Y},\sigma \right) \rightarrow C$ or simply $\underline{\mathcal{Y}}$%
. 
\end{definition}

Throughout the paper, all categories are assumed to be triangulated and dg. Note that one should allow singularities for the flat family of categories,
or $\Pi $-stability conditions, and nontrivial perverse conditions should be
needed to address these issues, which however will be neglected in this
article. $\Pi $-stability conditions of $\mathcal{Y}_{c}$ would be certain
Bridgeland stability conditions of the category $\mathcal{Y}_{c}$, which we
do not have a definition in general. When $\mathcal{Y}_{c}=Fuk\left(
Y_{c},\varpi _{c}\right) $ (resp. $ D^{b}\left( Y_{c},\Omega _{c}\right) $), a $\Pi $-stability condition of $\mathcal{Y}_{c}
$ would be a stability condition whose central charge is given by a holomorphic volume form $\Omega
_{c}$ (resp. a complexified K\"ahler class $ \varpi_{c} $) of $Y_{c}$. We simply denote the $\Pi $-stability condition by $\Omega_{c}$ (resp. $ \varpi_{c} $).  The stability condition $ \varpi_{c} $ is conjectured to exist near large volume limit point. 
In previous definitions (for instance, in \cite{KRS}) of A-branes and B-branes,  $\Pi $-stability conditions
were not included. We will simply call a physical 3d brane a 3d brane for
the rest of this article. The following table summarizes these conditions.%
\begin{equation*}
\begin{tabular}{c||c|c}
$\left\{ \left( \mathcal{Y}_{c},\sigma _{c}\right) \right\} _{c\in C}$ & $%
\text{3d A-brane}$ & $\text{3d B-brane}$ \\ \hline\hline
$C$ & $\text{cpx Lagr}$ & $\text{cpx Lagr}$ \\ 
$\left\{ \mathcal{Y}_{c}\right\} _{c\in C}$ & $\text{flat}$ & $\text{%
holomorphic}$ \\ 
$\left\{ \sigma _{c}\right\} _{c\in C}$ & $\text{holomorphic}$ & $\text{flat}
$%
\end{tabular}%
\end{equation*}

\bigskip 

Let us elaborate further on the holomorphic/flat condition of the $\Pi $%
-stability conditions. Given a 3d A/B-brane $\left\{ \left( \mathcal{Y}%
_{c},\sigma _{c}\right) \right\} _{c\in C}$ on $X$. In A-model, $\left\{ 
\mathcal{Y}_{c}\right\} _{c\in C}$ is a flat family of categories. Locally,
they are all quasi-equivalent to a fixed triangulated dg-category, say $\mathcal{Y}_{0}$,
canonically. So locally we have a map $\sigma :C\rightarrow Stab\left( 
\mathcal{Y}_{0}\right) $ to the space of Bridgeland stability conditions and
the A-brane condition requires that $\sigma $ is a holomorphic map with
respect to the natural complex structure on $Stab\left( \mathcal{Y}%
_{0}\right) $ defined by Bridgeland \cite{Bridgeland}.

In B-model, $\left\{ \mathcal{Y}_{c}\right\} _{c\in C}$ is a holomorphic
family of categories, thus their K-groups $K\left( \mathcal{Y}_{c}\right) $%
's are locally constant over $C_{sm}$. The central charge of $\sigma _{c}$
is a homomorphism $Z_{\sigma _{c}}:K\left( \mathcal{Y}_{c}\right)
\rightarrow \mathbb{C}$ and the B-brane condition requires that $Z_{\sigma
_{c}}$ is locally constant in $c\in C_{sm}$. This is because a stability
condition is determined by its central charge, locally in $Stab\left( 
\mathcal{Y}_{0}\right) $ \cite{Bridgeland}.

\begin{remark}
For a quasi-symmetric toric CYn manifold $Y$, there is an explicit
construction of a stringy K\"{a}hler moduli space \cite{Spenko Van Den Bergh}
which is conjectured to be equals to $\mathcal{M}_{sympl}\left( Y\right) $.
In general, $\mathcal{M}_{sympl}\left( Y\right) $ should sit inside $%
Stab\left( D^{b}\left( Y,\Omega \right) \right) $.
\end{remark}

\bigskip

Given two 3d A/B-branes $\underline{\mathcal{Y}_{1}}=\left\{ \left( \mathcal{%
Y}_{1,c},\sigma _{1,c}\right) \right\} _{c\in C_1}$ and $\underline{\mathcal{Y}%
_{2}}=\left\{ \left( \mathcal{Y}_{2,c},\sigma _{2,c}\right) \right\} _{c\in
C_2}$ on $X$, there should be an "intersection theory" of $\underline{\mathcal{%
Y}_{1}}$ and $\underline{\mathcal{Y}_{2}}$ which produces a category $%
_{A/B}Hom_{X}\left( \underline{\mathcal{Y}_{1}},\underline{\mathcal{Y}_{2}}%
\right) $. It is conjectured that such structures give a 2-category of 3d
boundary conditions in A/B-model and homological 3d mirror symmetry
conjecture predicts that there are 3d mirror pairs of holomorphic symplectic
manifolds/stacks $X$ and $X^{!}$ so that their A-model 2-category and
B-model 2-category got interchanged with each other. This is verified for
the hypertoric case by Gammage and Hilburn in \cite{Gammage Hilburn}.

Let $ \pi_1: Y_1 \to C_1 $ and $ \pi_2:Y_2\to C_2 $ be two CY-fibrations such that $C_{1}$ and $C_{2}$ intersect transversely at a single point $p\in X$. We have 3d A/B-branes
\begin{eqnarray*}
	\underline{{}_A\mathcal{Y}_1}=\{ (Fuk(Y_{1,c},\varpi_{1,c}), \Omega_{1,c})\}_{c\in C_1}\\
	\underline{{}_B\mathcal{Y}_1}=\{ (D^b(Y_{1,c},\Omega_{1,c}), \varpi_{1,c})\}_{c\in C_1}
\end{eqnarray*}
from $ \pi_1 $, and 3d A/B-branes
\begin{eqnarray*}
	\underline{{}_A\mathcal{Y}_2}=\{ (Fuk(Y_{2,c},\varpi_{2,c}), \Omega_{2,c})\}_{c\in C_2}\\
	\underline{{}_B\mathcal{Y}_2}=\{ (D^b(Y_{2,c},\Omega_{2,c}), \varpi_{2,c})\}_{c\in C_2}
\end{eqnarray*}
from $ \pi_2 $. Since $Y_1\times_X Y_2=Y_{1,p}\times Y_{2,p} $ , we expect that
\begin{eqnarray*}
	_{A}Hom_{X}\left( \underline{{}_A\mathcal{Y}_{1}},\underline{{}_A\mathcal{Y}_{2}}%
	\right)  &=&Fuk\left( Y_{1,p}\times Y_{2,p}\right) \text{,} \\
	_{B}Hom_{X}\left( \underline{{}_B\mathcal{Y}_{1}},\underline{{}_B\mathcal{Y}_{2}}%
	\right)  &=&D^{b}\left( Y_{1,p}\times Y_{2,p}\right) \text{,}
\end{eqnarray*}%
at least in the first approximation. If $C_{1}$ and $C_{2}$ do not intersect
transversely, their morphisms are harder to define (see eg \cite{KRS}\cite%
{Pascaleff}).

3d mirror symmetry is a conjectural duality of 3d A-models and 3d B-models
between $X$ and $X^{!}$. There are many exciting recent discoveries on 3d
mirror symmetry (see. eg \cite{BFN}\cite{BLPW}\cite{Chan Leung 3d}).
Examples of 3d mirror pairs includes (i) Gale dual hypertoric varieties and
(ii) Nakajima quiver varieties and BFN spaces \cite{BFN}.

\bigskip

In \cite{Chan Leung 3d}\cite{Chan Leung}, the first author discovered an
intimate relationship between 3d mirror symmetry and mirror symmetry. Very
loosely speaking, if $X$ and $X^{!}$ is a pair of 3d mirror manifolds and $%
\underline{\mathcal{Y}_{i}}$ and $\underline{\mathcal{Y}_{i}}^{!}$ are
mirror 3d branes in $X$ and $X^{!}$ for $i=1,2$, then 
\begin{equation*}
_{A}Hom_{X}\left( \underline{\mathcal{Y}_{1}},\underline{\mathcal{Y}_{2}}%
\right) \simeq _{~B}Hom_{X^{!}}\left( \underline{\mathcal{Y}_{1}^{!}},%
\underline{\mathcal{Y}_{2}^{!}}\right) .
\end{equation*}%
Thus $Y_{1}^{-}\times _{X}Y_{2}$ and $Y_{1}^{!-}\times _{X^{!}}Y_{2}^{!}$
should be 2d mirror to each other. By reversing these arguments, Chan and
the first authors \cite{Chan Leung SYZ} proposed a SYZ construction of 3d
mirror manifold $X^{!}$ to $X$ as the (stacky) cotangent bundle of the
moduli space of rank one 3d A-branes in $X$. That is $X^{!}=T^{\ast }C^{!}$
with $C^{!}$ a moduli space 3d A-branes $\underline{\mathcal{Y}}=\left( C,%
\mathcal{Y},\sigma \right) $ in $X$ where $C$ is a complex Lagrangian
submanifold in $X$ and $\mathcal{Y}\rightarrow C$ is a rank one perverse
schober. Namely a flat family of categories over $C_{sm}\subset C$ with
constraint singular behavior along $C\backslash C_{sm}$ and the generic
fiber is $Vect_{\mathbb{C}}$, the category of complex vector spaces (with
trivial stability condition). Note that $Vect_{\mathbb{C}}\simeq Fuk\left(
pt\right) \simeq D^{b}\left( pt\right) $. In particular, for the 3d A-brane $%
\underline{\mathcal{Y}}$ in $X$, its mirror 3d B-brane in $X^{!}=T^{\ast
}C^{!}$ is $T_{\underline{\mathcal{Y}}}^{\ast }C^{!}$, the cotangent fiber
at $\left[ \underline{\mathcal{Y}}\right] \in C^{!}$ coupled with the
trivial family of category $Vect_{\mathbb{C}}$ with trivial $\Pi $-stability
condition over it.

When $X$ is a hypertoric stack, i.e. $X=T^{\ast }\left[ \mathbb{C}^{n}/%
\mathbb{C}^{\times k}\right] $, the moduli space of rank one perverse
schobers over $\left[ \mathbb{C}^{n}/\mathbb{C}^{\times k}\right] $ can be
naturally identified with $\left[ \mathbb{C}^{\ast n}/\mathbb{C}^{\ast
\times \left( n-k\right) }\right] $, the Gale dual toric stack to $\left[ 
\mathbb{C}^{n}/\mathbb{C}^{\times k}\right] $. Thus its cotangent stack $%
X^{!}=T^{\ast }\left[ \mathbb{C}^{\ast n}/\mathbb{C}^{\ast \times \left(
n-k\right) }\right] $ is the 3d mirror to $X$ \cite{Chan Leung SYZ}.

\section{Mirror symmetry in 3d in 3d mirror symmetry}

Given a CY3 manifold $Y$ we consider 3d $N=4$ $\sigma $-models on hyperk\"{a}%
hler manifolds $\mathcal{J}_{sympl}\left( Y\right) $ and $\mathcal{J}%
_{cpx}\left( Y\right) $, universal intermediate Jacobians in A-model and
B-model for $Y$. To simplify our discussions, we ignore the lattice
subbundles and simply look at%
\begin{eqnarray*}
X &=&T^{\ast }\mathcal{M}_{sympl}\left( Y\right) \text{,} \\
X^{!} &=&T^{\ast }\mathcal{M}_{cpx}\left( Y\right) \text{.}
\end{eqnarray*}%
We will explain that $X$ and $X^{!}$ exhibit \textit{certain} 3d mirror
behaviors. The closed string mirror symmetry between $ Y $ and $ Y^\vee $ predicts that
\begin{eqnarray*}
	X &=&T^{\ast }\mathcal{M}_{cpx}\left( Y^\vee\right) \text{,} \\
	X^{!} &=&T^{\ast }\mathcal{M}_{sympl}\left( Y^\vee\right) \text{.}
\end{eqnarray*}%
By no means we claim that $X$ and $X^{!}$ are 3d mirror to each
other in full generality. Note that most of the discussion in this section
works for CYn for any n.

\bigskip

Given any $\Omega \in \mathcal{M}_{cpx}\left( Y\right) $, we define a 3d
A-brane $\underline{_{A}\mathcal{C}_{\Omega }}$ in $X=T^{\ast }\mathcal{M}%
_{sympl}\left( Y\right) $, 
\begin{equation*}
\underline{_{A}\mathcal{C}_{\Omega }}=\left( \pi :\left( _{A}\mathcal{C}%
_{\Omega },\sigma _{_{A}\mathcal{C}_{\Omega }}\right) \rightarrow \mathcal{M}%
_{sympl}\left( Y\right) \right) 
\end{equation*}%
with $\pi ^{-1}\left( \varpi \right) \simeq D^{b}\left( Y,\Omega \right) $
equipped with $\Pi $-stability condition $\varpi $. Forgetting the $\Pi $%
-stability conditions at the moment, this local system of derived categories
over the stringy K\"{a}hler moduli space of $Y$, conjecturally equivalent to 
$\mathcal{M}_{sympl}\left( Y\right) $, was constructed and studied
extensively in algebraic geometry \cite{Spenko Van Den Bergh}\cite{Donovan
Wemyss}. For example the monodromy action $\pi _{1}\left( \mathcal{M}%
_{sympl}\left( Y\right) \right) \rightarrow Aut\left( D^{b}\left( Y,\Omega
\right) \right) $ carries nontrivial information of the derived category $%
D^{b}\left( Y,\Omega \right) $. It was also showed that the local system
extends to be a perverse schober on a partial compactification $\overline{%
\mathcal{M}_{sympl}\left( Y\right) }\simeq \mathbb{C}^{N}$ for quasi-symmetric CY3 $ Y $. It is clear that 
$D^{b}\left( Y,\Omega \right) $ is locally constant over $\mathcal{M}%
_{sympl}\left( Y\right) $ and its stability condition $\varpi $ depends
holomorphically in $\varpi \in \mathcal{M}_{sympl}\left( Y\right) .$ 

\bigskip 

Suppose that all deformations of $\underline{_{A}\mathcal{C}_{\Omega }}$ as
3d A-branes are of the same form, i.e. they depend only on $\Omega \in 
\mathcal{M}_{cpx}\left( Y\right) $, then $\mathcal{M}_{cpx}\left( Y\right) $
is the moduli space of such 3d A-branes in $X$. Equivalently, $X^{!}$ is the
cotangent bundle of a moduli space of 3d A-branes in $X$. Namely this is a
3d SYZ type construction of $X^{!}$ from $X$.

Thus the 3d A-brane $\underline{_{A}\mathcal{C}_{\Omega }}$ in $X$
corresponds the cotangent fiber at $\Omega $ in $T^{\ast }\mathcal{M}%
_{cpx}\left( Y\right) =X^{!}$. This give the 3d B-brane 
\begin{equation*}
\underline{_{B}\mathcal{C}_{\Omega }^{!}\ }=\left( \pi :\underline{Vect_{%
\mathbb{C}}}\rightarrow T_{\Omega }^{\ast }\mathcal{M}_{cpx}\left( Y\right)
\right) 
\end{equation*}%
in $X^{!}$. Here $\underline{Vect_{\mathbb{C}}}$ is the trivial family of
the category $Vect_{\mathbb{C}}\simeq D^{b}\left( pt\right) $ of vector
spaces with its canonical stability condition. We call this the cotangent
fiber 3d B-brane at $\Omega $ in $X^{!}$.

\bigskip 

As $\underline{Vect_{\mathbb{C}}}$ is a trivial family of the category $%
Vect_{\mathbb{C}}$ with trivial stability condition, it is also a 3d A-brane
over any complex Lagrangian submanifold. Next we consider the cotangent
fiber 3d A-brane $\underline{_{A}\mathcal{D}_{\varpi }}$ at $\varpi \in 
\mathcal{M}_{sympl}\left( Y\right) $ in $X$, i.e.   
\begin{equation*}
\underline{_{A}\mathcal{D}_{\varpi }}=\left( \pi :\underline{Vect_{\mathbb{C}%
}}\rightarrow T_{\varpi }^{\ast }\mathcal{M}_{sympl}\left( Y\right) \right) 
\end{equation*}

Is there a 3d \textit{mirror }B-brane in\textit{\ }$X^{!}$\textit{\ } to $%
\underline{_{A}\mathcal{D}_{\varpi }}$? We define a 3d B-brane on $X^{!}$ as
follow: 
\begin{equation*}
\underline{_{B}\mathcal{D}_{\varpi }^{!}}=\left( \pi :\left( _{B}\mathcal{D}%
_{\varpi }^{!},\sigma _{_{B}\mathcal{D}_{\varpi }^{!}}\right) \rightarrow 
\mathcal{M}_{cpx}\left( Y\right) \right) 
\end{equation*}%
with its fiber over $\Omega \in \mathcal{M}_{cpx}\left( Y\right) $ as the
derived category $D^{b}\left( Y,\Omega \right) $ equipped with stability
condition $\varpi $. It is clear that $D^{b}\left( Y,\Omega \right) $
depends holomorphically in $\Omega $ and our choice of stability condition $%
\varpi $ is fixed. Thus it is a 3d B-brane in $X^{!}$.

\bigskip 

Given any $\Omega \in \mathcal{M}_{cpx}\left( Y\right) $ and $\varpi \in 
\mathcal{M}_{sympl}\left( Y\right) $, the support of 3d A-branes $\underline{%
_{A}\mathcal{C}_{\Omega }}$ and $\underline{_{A}\mathcal{D}_{\varpi }}$ in $X
$ are $\mathcal{M}_{sympl}\left( Y\right) $ and $T_{\varpi }^{\ast }\mathcal{%
M}_{sympl}\left( Y\right) $ respectively. In particular, they intersect
transversely at a single point $\varpi \in X$, the categories over $\varpi
\in X$ are $D^{b}\left( Y,\Omega \right) $ and $Vect_{\mathbb{C}}$
respectively, with $\Pi $-stability conditions $\varpi $ and trivial
respectively. Thus 
\begin{equation*}
_{A}Hom_{X}\left( \underline{_{A}\mathcal{C}_{\Omega }},\underline{_{A}%
\mathcal{D}_{\varpi }}\right) \simeq D^{b}\left( Y,\Omega \right) 
\end{equation*}%
equipped with $\Pi $-stability condition $\varpi $.

Similarly, the support of 3d B-branes $\underline{_{B}\mathcal{C}_{\Omega }^{!}}$
and $\underline{_{B}\mathcal{D}_{\varpi }^{!}}$ in $X^{!}$ are $T_{\Omega
}^{\ast }\mathcal{M}_{cpx}\left( Y\right) $ and $\mathcal{M}_{cpx}\left(
Y\right) $ respectively, which intersect transversely at a single point $%
\Omega \in X^{!}$. Thus 
\begin{equation*}
_{B}Hom_{X}\left( \underline{_{B}\mathcal{C}_{\Omega }^{!}},\underline{_{B}%
\mathcal{D}_{\varpi }^{!}}\right) \simeq D^{b}\left( Y,\Omega \right)
\end{equation*}%
equipped with $\Pi $-stability condition $\varpi $ as well.

The 3d mirror phenomena for $X=T^{\ast }\mathcal{M}_{sympl}\left( Y\right) $
and $X^{!}=T^{\ast }\mathcal{M}_{cpx}\left( Y\right) $ are (i) realizing $%
X^{!}$ as the cotangent bundle of a moduli space of 3d A-branes $\underline{%
_{A}\mathcal{C}_{\Omega }}$'s in $X$ and (ii) identifying 3d Hom categories 
\begin{equation*}
_{A}Hom_{X}\left( \underline{_{A}\mathcal{C}_{\Omega }},\underline{_{A}%
\mathcal{D}_{\varpi }}\right) \simeq \text{ }_{B}Hom_{X}\left( \underline{%
_{B}\mathcal{C}_{\Omega }^{!}},\underline{_{B}\mathcal{D}_{\varpi }^{!}}\right) 
\end{equation*}%
with $\Pi $-stability conditions for any $\Omega \in \mathcal{M}_{cpx}\left(
Y\right) $ and $\varpi \in \mathcal{M}_{sympl}\left( Y\right) $. We could
also view (ii) as a construction of $\underline{_{B}\mathcal{D}_{\varpi }^{!}}$
via a 3d SYZ type transformation of the cotangent fiber brane $\underline{%
_{A}\mathcal{D}_{\varpi }}$ as $\mathcal{M}_{cpx}\left( Y\right) $
parametrizes $\underline{_{A}\mathcal{C}_{\Omega }}$'s in $X$.

\bigskip\ 

Similarly, we could also use the Fukaya category $Fuk\left( Y,\varpi \right) 
$ with $\Pi $-stability condition $\Omega $ to construct 3d B-branes on $X$,%
\begin{equation*}
\underline{_{B}\mathcal{E}_{\Omega }}=\left( \pi :\left( _{B}\mathcal{E}%
_{\Omega },\sigma _{_{B}\mathcal{E}_{\Omega }}\right) \rightarrow \mathcal{M}%
_{sympl}\left( Y\right) \right) ,
\end{equation*}%
with its fiber over $\varpi \in \mathcal{M}_{sympl}\left( Y\right) $ as $%
\left( Fuk\left( Y,\varpi \right) ,\Omega \right) $, i.e. the Fukaya category $%
Fuk\left( Y,\varpi \right) $ with $\Pi $-stability condition $\Omega $; and
to construct 3d A-branes on $X^{!}$%
\begin{equation*}
\underline{_{A}\mathcal{F}_{\varpi }^{!}}=\left( \pi :\left( _{A}\mathcal{F}%
_{\varpi },\sigma _{_{A}\mathcal{F}_{\varpi }}\right) \rightarrow \mathcal{M}%
_{cpx}\left( Y\right) \right) ,
\end{equation*}%
with its fiber over $\Omega \in \mathcal{M}_{cpx}\left( Y\right) $ as $%
\left( Fuk\left( Y,\varpi \right) ,\Omega \right) $. Their corresponding 3d
mirror branes are cotangent fiber 3d A-brane $\underline{_{A}\mathcal{E}%
_{\Omega }^{!}}$ at $\Omega $ in $X^{!}$ and cotangent fiber 3d B-brane $%
\underline{_{B}\mathcal{F}_{\varpi }}$ at $\varpi $ in $X$.

\bigskip 

Let us summarize these structures in the following table

\begin{equation*}
\begin{tabular}{cc}
$X=T^{\ast }\mathcal{M}_{sympl}\left( Y\right) $ & $X^{!}=T^{\ast }\mathcal{M%
}_{cpx}\left( Y\right) $ \\ 
\begin{tabular}{l||l|l|l}
& $A/B$ & $C$ & $\text{fiber}$ \\ \hline\hline
$\underline{_{A}\mathcal{C}_{\Omega }}$ & $A$ & $\mathcal{M}_{sympl}\left(
Y\right) $ & $D^{b}\left( Y,\Omega \right) $ \\ 
$\underline{_{A}\mathcal{D}_{\varpi }}$ & $A$ & $T_{\varpi }^{\ast }$ & $%
Vect_{\mathbb{C}}$ \\ 
$\underline{_{B}\mathcal{E}_{\Omega }}$ & $B$ & $\mathcal{M}_{sympl}\left(
Y\right) $ & $Fuk\left( Y,\varpi \right) $ \\ 
$\underline{_{B}\mathcal{F}_{\varpi }}$ & $B$ & $T_{\varpi }^{\ast }$ & $%
Vect_{\mathbb{C}}$%
\end{tabular}
& 
\begin{tabular}{l||l|l|l}
 & $A/B$  & $C$ & $\text{fiber}$ \\ \hline\hline
$\underline{_{B}\mathcal{C}_{\Omega }^{!}}$ & $B$ & $T_{\Omega }^{\ast }$ & $%
Vect_{\mathbb{C}}$ \\ 
$\underline{_{B}\mathcal{D}_{\varpi }^{!}}$ & $B$ & $\mathcal{M}_{cpx}\left(
Y\right) $ & $D^{b}\left( Y,\Omega \right) $ \\ 
$\underline{_{A}\mathcal{E}_{\Omega }^{!}}$ & $A$ & $T_{\Omega }^{\ast }$ & $%
Vect_{\mathbb{C}}$ \\ 
$\underline{_{A}\mathcal{F}_{\varpi }^{!}}$ & $A$ & $\mathcal{M}_{cpx}\left(
Y\right) $ & $Fuk\left( Y,\varpi \right) $%
\end{tabular}%
\end{tabular}%
.
\end{equation*}%
Under mirror symmetry, the roles between $X$ and $X^{!}$ will get
interchanged if we replace $Y$ by its mirror CY $Y^{\vee }$ as follow,$%
\overset{}{}$%
\begin{eqnarray*}
\mathcal{M}_{sympl}\left( Y\right)  &\simeq &\mathcal{M}_{cpx}\left( Y^{\vee
}\right)  \\
\mathcal{M}_{cpx}\left( Y\right)  &\simeq &\mathcal{M}_{sympl}\left( Y^{\vee
}\right)  \\
\left( D^{b}\left( Y,\Omega \right) ,\varpi \right)  &\simeq &\left(
Fuk\left( Y^{\vee },\varpi ^{\vee }\right) ,\Omega ^{\vee }\right)  \\
\left( Fuk\left( Y,\varpi \right) ,\Omega \right)  &\simeq &\left(
D^{b}\left( Y^{^{\vee }},\Omega ^{^{\vee }}\right) ,\varpi ^{^{\vee
}}\right) .
\end{eqnarray*}

In the next two sections, we will construct new 3d A- and B-branes on $X$
and $X^{!}$.

\section{Quintic CY3}

In this section , we construct 3d branes in $ X $ and $ X^! $ whose
supports are neither fibers nor zero sections using symmetries of $ Y $. Throughout this section, $Y$ is a smooth quintic CY3 in $\mathbb{CP}^{4}$,
i.e. 
\begin{eqnarray*}
	Y=\left\{ f\left( z_{0},...,z_{4}\right) =0\right\} \subset \mathbb{CP}^{4}
\end{eqnarray*}
with $f$ a degree $5$ homogeneous polynomial in $\left(
z_{0},...,z_{4}\right) $. We have $h^{1,1}\left( Y\right) =1$ and $%
h^{2,1}\left( Y\right) =101$. That is $\dim \mathcal{M}_{sympl}\left(
Y\right) =2$ and $\dim \mathcal{M}_{cpx}\left( Y\right) =102$. In particular, $ h^{2,1}(Y^\vee)=1 $, where $ Y^\vee $ is the mirror to $ Y $. The one parameter family of 
$Y^\vee$ can be constructed explicitly as follow \cite{Greene Plesser}: Consider
a special one-parameter family of quintic CY3 $Y_{\psi }=\left\{ f_{\psi
}=0\right\} \subset \mathbb{CP}^{4}$ for $\psi \in \mathbb{C}$ with%
\begin{equation*}
f_{\psi }\left( z_{0},...,z_{4}\right)
=\sum_{j=0}^{4}z_{j}^{5}+\psi \prod_{j=0}^{4}z_{j}.
\end{equation*}%
Each $Y_{\psi }$ admits a discrete symmetry group $\Gamma \simeq \mathbb{Z}%
_{5}^{3}$ and the mirror to $Y$ is a CY resolution of $Y_{\psi }/\Gamma $,
that is 
\begin{equation*}
Y^{\vee }\simeq \widetilde{Y_{\psi }/\Gamma }\text{.}
\end{equation*}%
In particular, this gives a natural inclusion,%
\begin{equation*}
\mathcal{M}_{cpx}\left( Y^{\vee }\right) \subset \mathcal{M}_{cpx}\left(
Y\right) .
\end{equation*}

Concretely, the componentwise action by $c=\left( c_{0},\cdots ,c_{4}\right)
\in \left( \mathbb{Z}_{5}\right) ^{5}\subset \left( \mathbb{C}^{\times
}\right) ^{5}$ on $\mathbb{C}^{5}$ preserves $f_{\psi }\left(
z_{0},...,z_{4}\right) $ provided that $\prod\nolimits_{i}c_{i}=1$. The
diagonal action by $\mathbb{Z}_{5}\overset{\Delta }{\hookrightarrow }\left( 
\mathbb{Z}_{5}\right) ^{5}$ will be trivial upon projectification. The
resulting symmetry group $\Gamma $ is therefore given by the following exact
sequence of Abelian groups

\begin{equation*}
1\rightarrow \Gamma \rightarrow \mathbb{Z}_{5}^{5}/\Delta _{\mathbb{Z}_{5}}%
\overset{\Pi }{\rightarrow }\mathbb{Z}_{5}\rightarrow 1
\end{equation*}%
and satisfies $\Gamma \simeq \mathbb{Z}_{5}^{3}$. 

Closed string mirror symmetry conjecture says that 
\begin{equation*}
\mathcal{M}_{cpx}\left( Y^{\vee }\right) \simeq \mathcal{M}_{sympl}\left(
Y\right) \text{,}
\end{equation*}
which was described physically by Candelas et al \cite%
{CdGP} and proven mathematically by Givental \cite{Givental} and Lian, Liu
and Yau \cite{LLY}. Kontsevich's HMS conjecture \cite{Kontsevich ICM}%
\begin{equation*}
D^{b}\left( Y^\vee,\Omega^\vee \right) \simeq Fuk\left( Y,\varpi \right) 
\end{equation*}%
for quintic CY3 was proven by Sheridan \cite{Sheridan}. 

\bigskip 

Recall given any $\Omega \in \mathcal{M}_{cpx}\left( Y\right) $, $\underline{%
_{A}\mathcal{C}_{\Omega }}$ is a 3d A-brane in $T^{\ast }\mathcal{M}%
_{sympl}\left( Y\right) $ with fiber over $\varpi \in \mathcal{M}%
_{sympl}\left( Y\right) $ being quasi-equivariant to $D^{b}\left( Y,\Omega \right) $ with $\Pi $%
-stability condition $\varpi $. Via the natural inclusion $\mathcal{M}%
_{cpx}\left( Y^{\vee }\right) \subset \mathcal{M}_{cpx}\left( Y\right) $ and
the closed string mirror symmetry $\mathcal{M}_{sympl}\left( Y\right) \simeq 
\mathcal{M}_{cpx}\left( Y^{\vee }\right) $, we have an inclusion%
\begin{equation*}
\iota :\mathcal{M}_{sympl}\left( Y\right) \hookrightarrow \mathcal{M}%
_{cpx}\left( Y\right) .
\end{equation*}%
The conormal bundle of $ \mathrm{graph}(\iota)\subset  \mathcal{M}_{sympl}\left( Y\right) \times  \mathcal{M}%
_{cpx}\left( Y\right) $ is a complex
Lagrangian submanifold $\mathcal{K}\subset X\times X^{!}$. Via the
Lagrangian correspondence with kernel $\mathcal{K}$, the 3d A-brane  $\underline{_{A}\mathcal{%
C}_{\Omega }}$ in $ X $ induces a 3d A-brane in $X^{!}$ with support the conormal bundle $N_{\mathcal{M}%
_{sympl}\left( Y\right) /\mathcal{M}_{cpx}\left( Y\right) }^{\ast }$ and
fiber over $\varpi \in \mathcal{M}_{sympl}\left( Y\right) $ being quasi-equivalent to $%
D^{b}\left( Y,\Omega \right) $ with $\Pi $-stability condition $\varpi $. We
denote this 3d A-brane on $X^{!}$ as $\underline{_{A}\mathcal{C}_{\Gamma
,\Omega }^{!}}$.

The 3d A-branes $\underline{_{A}\mathcal{C}_{\Gamma ,\Omega }^{!}}$ and $%
\underline{_{A}\mathcal{F}_{\varpi }^{!}}$ intersect cleanly along $\mathcal{%
M}_{sympl}\left( Y\right) \subset \mathcal{M}_{cpx}\left( Y\right) $. The
fibers over any such intersection point $\varpi ^{\prime }\in \mathcal{M}%
_{sympl}\left( Y\right) $ are%
\begin{equation*}
\left( D^{b}\left( Y,\Omega \right) ,\varpi ^{\prime }\right) \text{ and }%
\left( Fuk\left( Y,\varpi \right) ,\Omega _{\varpi ^{\prime }}\right) 
\end{equation*}%
respectively. Here $\Omega _{\varpi ^{\prime }}=\iota \left( \varpi ^{\prime
}\right) $ is the element in corresponding to $\varpi ^{\prime }$ under the
inclusion $\iota :\mathcal{M}_{sympl}\left( Y\right) \subset \mathcal{M}%
_{cpx}\left( Y\right) $. Switching the A- and B-sides of the above two
situations, we have%
\begin{equation*}
\text{ }\left( Fuk\left( Y,\varpi ^{\prime }\right) ,\Omega \right) \text{
and }\left( D^{b}\left( Y,\Omega _{\varpi ^{\prime }}\right) ,\varpi \right) 
\text{.}
\end{equation*}%
By varying $\varpi ^{\prime }\in \mathcal{M}_{sympl}\left( Y\right) $, we
obtain two 3d B-branes on $X=T^{\ast }\mathcal{M}_{sympl}\left( Y\right) $,
and both have supports equal to $\mathcal{M}_{sympl}\left( Y\right) $.

\bigskip 

Similarly, (i) for any $\Omega \in \mathcal{M}_{cpx}\left( Y\right) $, we
can construct a 3d B-brane on $X^{!}$ with support $N_{\mathcal{M}%
_{sympl}\left( Y\right) /\mathcal{M}_{cpx}\left( Y\right) }^{\ast }$ and
fiber $Fuk\left( Y,\varpi ^{\prime }\right) $ over $\varpi ^{\prime }\in 
\mathcal{M}_{sympl}\left( Y\right) $ and with $\Pi $-stability condition $%
\Omega $; (ii) for any $\varpi \in \mathcal{M}_{sympl}\left( Y\right) $, we
can construct a 3d A-brane on $X$ with support $\mathcal{M}_{sympl}\left(
Y\right) $ and fiber $Fuk\left( Y,\varpi \right) $ over $\varpi ^{\prime
}\in \mathcal{M}_{sympl}\left( Y\right) $ and with $\Pi $-stability
condition $\Omega _{\varpi ^{\prime }}$. 

To summarize, given $\varpi \in \mathcal{M}_{sympl}\left( Y\right) $ we have
following 3d branes in $X=T^{\ast }\mathcal{M}_{sympl}\left( Y\right) $,%
\begin{equation*}
\begin{tabular}{c||c|c|c}
$A/B$ & Support & Fiber over $\varpi ^{\prime }$ & $\Pi $-stability condition
\\ \hline\hline
$A$ & $\mathcal{M}_{sympl}\left( Y\right) $ & $Fuk\left( Y,\varpi \right) $
& $\Omega _{\varpi ^{\prime }}$ \\ 
$B$ & $\mathcal{M}_{sympl}\left( Y\right) $ & $D^{b}\left( Y,\Omega _{\varpi
^{\prime }}\right) $ & $\varpi $%
\end{tabular}%
.
\end{equation*}%
Similarly, given $\Omega \in \mathcal{M}_{cpx}\left( Y\right) $ we have
following 3d branes in $X^{!}=T^{\ast }\mathcal{M}_{cpx}\left( Y\right) $,%
\begin{equation*}
\begin{tabular}{c||c|c|c}
$A/B$ & Support & Fiber over $\varpi ^{\prime }$ & $\Pi $-stability condition
\\ \hline\hline
$A$ & $N_{\mathcal{M}_{sympl}\left( Y\right) /\mathcal{M}_{cpx}\left(
Y\right) }^{\ast }$ & $D^{b}\left( Y,\Omega \right) $ & $\varpi ^{\prime }$
\\ 
$B$ & $N_{\mathcal{M}_{sympl}\left( Y\right) /\mathcal{M}_{cpx}\left(
Y\right) }^{\ast }$ & $Fuk\left( Y,\varpi ^{\prime }\right) $ & $\Omega $%
\end{tabular}%
.
\end{equation*}

\bigskip

This construction can be generalized to a class of CYn manifolds $Y$ with
finite Abelian symmetry group $\Gamma $ so that $Y^{\vee }\simeq \widetilde{%
Y/\Gamma }$. In general, there are a decreasing sequence of subgroups%
\begin{equation*}
\Gamma =\Gamma _{r}\supset \Gamma _{r-1}\supset \cdots \cdots \supset \Gamma
_{1}\supset \Gamma _{0}=\left\{ 0\right\} \text{,}
\end{equation*}%
so that the mirror to $\widetilde{Y/\Gamma _{j}}$ is $\widetilde{Y/\Gamma
_{r-j}}$. If we denote $Y_{j}=\widetilde{Y/\Gamma _{j}}$, then%
\begin{equation*}
\left( Y_{j}\right) ^{\vee }\simeq Y_{r-j}\text{.}
\end{equation*}%
By the mirror symmetry conjecture, we have%
\begin{equation*}
\begin{tabular}{lllllllll}
$\mathcal{M}_{cpx}\left( Y_{0}\right) $ & $\supset $ & $\mathcal{M}%
_{cpx}\left( Y_{1}\right) $ & $\supset $ & $\cdots $ & $\supset $ & $%
\mathcal{M}_{cpx}\left( Y_{j}\right) $ & $\supset $ & $\cdots $ \\ 
$\parallel $ &  & $\parallel $ &  &  &  & $\parallel $ &  &  \\ 
$\mathcal{M}_{sympl}\left( Y_{r}\right) $ &  & $\mathcal{M}_{sympl}\left(
Y_{r-1}\right) $ &  &  &  & $\mathcal{M}_{sympl}\left( Y_{r-j}\right) $ &  & 
\end{tabular}%
.
\end{equation*}%
The above construction produce new 3d branes on $X$ and $X^{!}$ as follow:\
For any $j$, given $\varpi \in \mathcal{M}_{sympl}\left( Y\right) $ we have
following 3d branes on $X$%
\begin{equation*}
\begin{tabular}{c||c|c|c}
$A/B$ & Support & Fiber over $\varpi ^{\prime }\in \mathcal{M}_{sympl}\left(
Y\right) $ & $\Pi $-stability condition \\ \hline\hline
$A$ & $\mathcal{C}$ & $Fuk\left( Y,\varpi \right) $ & $\Omega _{\varpi
^{\prime }}$ \\ 
$B$ & $\mathcal{M}_{sympl}\left( Y\right) $ & $D^{b}\left( Y_{j},\Omega
_{\varpi ^{\prime }}\right) $ & $\varpi $%
\end{tabular}%
,
\end{equation*}%
and given $\Omega \in \mathcal{M}_{cpx}\left( Y\right) $ we have following
3d branes on $X^{!}$ 
\begin{equation*}
\begin{tabular}{c||c|c|c}
$A/B$ & Support & Fiber over $\varpi ^{\prime }\in \mathcal{M}_{sympl}\left(
Y_{r-j}\right) $ & $\Pi $-stability condition \\ \hline\hline
$A$ & $N_{\mathcal{M}_{sympl}\left( Y_{r-j}\right) /\mathcal{M}_{cpx}\left(
Y\right) }^{\ast }$ & $D^{b}\left( Y,\Omega \right) $ & $\varpi ^{\prime }$
\\ 
$B$ & $N_{\mathcal{M}_{sympl}\left( Y_{r-j}\right) /\mathcal{M}_{cpx}\left(
Y\right) }^{\ast }$ & $Fuk\left( Y_{r-j},\varpi ^{\prime }\right) $ & $%
\Omega $%
\end{tabular}%
.
\end{equation*}%
Of course, for any $i$ we could obtain 3d A- and B-branes on each of the $%
T^{\ast }\mathcal{M}_{sympl}\left( Y_{i}\right) $ and $T^{\ast }\mathcal{M}%
_{cpx}\left( Y_{i}\right) $ in similar fashions.

\section{3d DT-branes}

When $\left( Y,\varpi ,\Omega \right) $ is a CY3 manifold. Donaldson and
Thomas \cite{DT} proposed an invariant $DT_{\left( Y,\Omega \right) ,\varpi }\left(
\gamma \right) \in \mathbb{Q}$ which counts $\varpi $-semi-stable coherent
sheaves $E$ on $\left( Y,\Omega \right) $ with $ch\left( E\right) =\gamma $. The proposal led to many subsequent works of its mathematical definition.
Under orientation datum type assumptions, there would have categorified
generalizations, the cohomology DT-invariant $H_{\left( Y,\Omega \right)
,\varpi }^{DT}\left( \gamma \right) \in Vect_{\mathbb{C}}$ and further
categorification DT-category $\mathcal{C}^{\mathrm{DT}}_{\left( Y,\Omega \right) ,\varpi
}\left( \gamma \right) $, so that $DT_{\left( Y,\Omega \right) ,\varpi
}\left( \gamma \right) =\dim H_{\left( Y,\Omega \right) ,\varpi }^{DT}\left(
\gamma \right) $ and $H_{\left( Y,\Omega \right) ,\varpi }^{DT}\left( \gamma
\right) =HH^{\ast }\left( \mathcal{C}^{\mathrm{DT}}_{\left( Y,\Omega \right) ,\varpi
}\left( \gamma \right) \right) $, the Hochschild cohomology of DT-category. 

We conjecture that there is a 3d B-brane $_{B}\underline{\mathcal{DT}}%
_{\varpi }\left( \gamma \right) $ on $X^{!}$ so that 
\begin{equation*}
\mathcal{C}^{\mathrm{DT}}_{\left( Y,\Omega \right) ,\varpi }\left( \gamma \right) =\text{~%
}_{B}Hom_{X}\left( _{B}\underline{\mathcal{DT}}_{\varpi }\left( \gamma
\right) ,\underline{_{B}\mathcal{C}_{\Omega }^{!}}\right) .
\end{equation*}

To describe $_{B}\underline{\mathcal{DT}}_{\varpi }\left( \gamma \right) $
when $E$ is rank $r$ Hermitian vector bundle over $Y$, we consider
the universal Chern-Simons (abbrev. CS) functional,

\begin{eqnarray*}
CS &:&\mathcal{M}_{cpx}\left( Y\right) \times \mathcal{A}\left( E\right)
\rightarrow \mathbb{C} \\
CS\left( \Omega ,A\right)  &=&\int_{Y}\Omega \wedge Tr_{E}\left( A\wedge dA+%
\frac{2}{3}A^{3}\right)  \\
&=&const.\int_{Y\times \left[ 0,1\right] _{t}}\Omega \wedge Tr_{E}F_{\mathbb{%
A}}^{2}
\end{eqnarray*}%
where $d+\mathbb{A}=d_{Y}+\frac{d}{dt}dt+tA$ is a connection on the pullback
bundle of $E$ over $Y\times \left[ 0,1\right] _{t}\rightarrow Y$ joining $d$
and $d+A$ on $Y$. $\mathcal{A}\left( E\right) $ is the space of Hermitian
connections on $E$.

If we restrict $CS$ to $\left\{ \Omega \right\} \times $ $\mathcal{A}\left(
E\right) $, then the Euler-Lagrange equation is 
\begin{equation*}
\Omega \wedge F=0\in \Omega ^{5}\left( Y,\mathbb{C}\right) \text{,}
\end{equation*}%
which is equivalent to $F^{0,2}=0$, with respect to the complex structure determined by $ \Omega $. Such a connection $A$ defines a
holomorphic structure on $E$ over $\left( Y,\Omega \right) $. 

Graph of the differential of\ CS\ is a complex Lagrangian submanifold in $%
T^{\ast }\left( \mathcal{M}_{cpx}\left( Y\right) \times \mathcal{A}\left(
E\right) \right) =X^{!}\times T^{\ast }\mathcal{A}\left( E\right) $,%
\begin{equation*}
Gr\left( dCS\right) \subset X^{!}\times T^{\ast }\mathcal{A}\left( E\right) 
\text{.}
\end{equation*}%
Consider the product of the diagonal Lagrangian $X^{!}\subset X^{!-}\times
X^{!}$ and the zero section Lagrangian $\mathcal{A}\left( E\right) \subset
T^{\ast }\mathcal{A}\left( E\right) $, we have a complex Lagrangian
submanifold $\mathcal{L}$ 
\begin{equation*}
\mathcal{L}=X^{!}\times \mathcal{A}\left( E\right) \subset X^{!-}\times X^{!}\times 
T^*\mathcal{A}\left( E\right) \text{.}
\end{equation*}%
This gives a complex Lagrangian correspondence $\Phi _{\mathcal{L}}$ from $%
X^{!}\times T^* \mathcal{A}\left( E\right) $ to $X^{!}$. Under clean
intersection assumption, 
\begin{equation*}
\Phi _{\mathcal{L}}\left( Gr\left( dCS\right) \right) \subset X^{!}
\end{equation*}%
is a complex Lagrangian submanifold which should give a 3d B-branes in $X^{!}
$. However we need to quotient out the $\mathcal{G}_{\mathbb{C}}$-symmetry
which requires us to pick a stability condition $\varpi \in \mathcal{M}%
_{sympl}\left( Y\right) $ in order to obtain the 3d DT B-brane $_{B}%
\underline{\mathcal{DT}}_{\varpi }\left( \gamma \right) $. There are many
nontrivial issues needed to be resolved to make this construction rigorous.
For instance it is not clear how to put a $\Pi $-stability condition for
fiber categories in $_{B}\underline{\mathcal{DT}}_{\varpi }\left( \gamma
\right) $.

Recall that the support of $\underline{\mathcal{C}_{\Omega }^{!}}$ is $%
T_{\Omega }^{\ast }\mathcal{M}_{cpx}\left( Y\right) $. Note that
intersection points of $_{B}\underline{\mathcal{DT}}_{\varpi }\left( \gamma
\right) $ and $T_{\Omega }^{\ast }\mathcal{M}_{cpx}\left( Y\right) $
correspond to holomorphic structures on $E$ over $\left( Y,\Omega \right) $.
Thus their intersection number, counted with multiplicities, should be the
DT invariants, and their categorical intersections would give the DT-categories.

\begin{remark}
Given any $\Omega \in \mathcal{M}_{cpx}\left( Y\right) $, one is tempted to
define a 3d A-brane on $X=T^{\ast }\mathcal{M}_{sympl}\left( Y\right) $ with
support $\mathcal{M}_{sympl}\left( Y\right) $ and fiber over $\varpi \in 
\mathcal{M}_{sympl}\left( Y\right) $ being the DT-category $\mathcal{C}^{\mathrm{DT}}%
_{\left( Y,\Omega \right) ,\varpi }\left( \gamma \right) $. However
DT-invariant can jump as $\varpi $ crosses real codimension one walls in $%
\mathcal{M}_{sympl}\left( Y\right) $ \cite{KS}. We do not know how to
resolve this issue. As the Wall Crossing Formula (WCF) relates DT invariants
for all $\gamma $'s together, we might consider to combine $\mathcal{C}^{\mathrm{DT}}%
_{\left( Y,\Omega \right) ,\varpi }\left( \gamma \right) $'s for all $\gamma 
$'s together.
\end{remark}

\bigskip

Similarly we could use the analog of DT-theory for the counting of special
Lagrangian submanifolds \cite{Joyce} in $\left( Y,\varpi ,\Omega \right) $
to construct 3d B-branes in $X$ as follow: Consider a fixed three manifold $M
$ with a Hermitian line bundle $E$ over $M$. We define the A-sided analog of
the Chern-Simons functional as below:%
\begin{eqnarray*}
_{A}CS &:&\mathcal{M}_{sympl}\left( Y\right) \times Map\left( M,Y\right)
\times \mathcal{A}\left( E\right) \rightarrow \mathbb{C} \\
_{A}CS\left( \varpi ,u,A\right)  &=&\ \int_{M\times \left[ 0,1\right] _{t}}%
\mathbf{u}^{\ast }\varpi \wedge e^{F_{\mathbb{A}}} \\
&=&t\int_{M\times \left[ 0,1\right] _{t}}\exp \left( \mathbf{u}^{\ast
}\omega +\left(i \mathbf{u}^{\ast }B+F_{\mathbb{A}}\right) \right) .
\end{eqnarray*}%
Here $\varpi =t\exp \left( \omega +iB\right) $, $\mathbb{A}$ is the
Hermitian connection over $M\times \left[ 0,1\right] _{t}$ connecting $d+A$
and $d$ as before, $\mathbf{u}$ is a map from $M\times \left[ 0,1\right] _{t}
$ to $Y$ connecting $u$ and a fixed background map $u_{0}:M\rightarrow Y$.

When we restrict $_{A}CS$ to $\left\{ \varpi \right\} \times $ $Map\left(
M,Y\right) \times \mathcal{A}\left( E\right) $, the Euler-Lagrange
equations are 
\begin{equation*}
u^{\ast }\omega =0\text{ and }iu^{\ast }B+F_{A}=0\text{. }
\end{equation*}%
Namely it is a 2d A-brane in $\left( Y,\varpi \right) $.

To count these solutions, we need to quotient out $\mathcal{G}$, the group
of unitary gauge transformations of $E$ and $C^{\infty }\left(
M,S^{1}\right) $, the group of Hamiltonian symmetries, or instead we could
impose \textit{special Lagrangian} condition 
\begin{equation*}
u^{\ast }\func{Re}\Omega =0\text{.}
\end{equation*}%
Modulo difficult analytic issues, this defines $_{A}DT_{\left(
Y,\varpi \right) ,\Omega }\left( \gamma \right) \in \mathbb{Q}$ which counts
special Lagrangian cycles in $Y$ representing the class $\gamma =\left[
u\left( M\right) \right] \in H_{3}\left( Y,\mathbb{Z}\right) $.

Similarly $\Phi _{_{A}\mathcal{L}}\left( Gr\left( d_{A}CS\right) \right) $
defines a complex Lagrangian in $X$, thus a B-brane $_{B}\underline{\mathcal{%
DT}}_{\Omega }^{SLag}\left( \gamma \right) $ in $X$. Again 
\begin{equation*}
DT_{\left( Y,\varpi \right) ,\Omega }^{SLag}\left( \gamma \right) =\#\left(
_{B}\underline{\mathcal{DT}}_{\Omega }^{SLag}\left( \gamma \right) \cap
T_{\varpi }^{\ast }\mathcal{M}_{sympl}\left( Y\right) \right) 
\end{equation*}%
should be the A-side analog of DT invariants which count the number of
special Lagrangian cycles in $Y$ representing the class $\gamma $, and 
\begin{equation*}
H_{\left( Y,\varpi \right) ,\Omega }^{DT,SLag}\left( \gamma \right)
=HF\left( _{B}\underline{\mathcal{DT}}_{\Omega }^{SLag}\left( \gamma \right)
,T_{\varpi }^{\ast }\mathcal{M}_{sympl}\left( Y\right) \right) 
\end{equation*}%
should be its categorification, and $_{B}Hom_{X}\left( _{B}\underline{%
\mathcal{DT}}_{\Omega }^{SLag}\left( \gamma \right) ,\underline{_{B}\mathcal{%
F}_{\varpi }}\right) $ would be its further categorification.

\bigskip 

If we look at the 3d A-side on the universal intermediate Jacobian $\mathcal{J}_{sympl}\left( Y\right) =T_{%
\mathbb{Z}}^{\ast }\mathcal{M}_{sympl}\left( Y\right) \backslash T^{\ast }%
\mathcal{M}_{sympl}\left( Y\right) $, which is an Abelian varieties
fibration over $\mathcal{M}_{sympl}\left( Y\right) $ with a complex
Lagrangian section $S$. We denote its fiber over $\varpi \in \mathcal{M}%
_{sympl}\left( Y\right) $ as $F_{\varpi }$. Even though $S$ and $F_{\varpi }$
intersect transversely at a single point $\varpi \in \mathcal{J}%
_{sympl}\left( Y\right) $, the A-model Hom category $_{A}Hom_{X}\left(
S,F_{\varpi }\right) $ is not simply $Vect_{\mathbb{C}}$ as $H_{1}\left(
F_{\varpi },\mathbb{Z}\right) \simeq H^{3}\left( Y,\mathbb{Z}\right) \neq 0$%
. Formally, $_{A}Hom_{X}\left( S,F_{\varpi }\right) $ is the category
defined by the complex Morse theory \cite{GMW} of the complex symplectic area functional $f$ on the loop space of $%
\mathcal{J}_{sympl}\left( Y\right) $,%
\begin{equation*}
f:\widetilde{\Omega }_{S\rightarrow F_{\varpi }}\left( \mathcal{J}%
_{sympl}\left( Y\right) \right) \rightarrow \mathbb{C}.
\end{equation*}%
Here $\Omega _{S\rightarrow F_{\varpi }}\left( \mathcal{J}_{sympl}\left(
Y\right) \right) $ is the space of all paths $u$ in $\mathcal{J}%
_{sympl}\left( Y\right) $ from $S$ to $F_{\varpi }$ and $\widetilde{\Omega }%
_{S\rightarrow F_{\varpi }}\left( \mathcal{J}_{sympl}\left( Y\right) \right) 
$ parametrizes $u$'s together with a certain path $\mathbf{u}$ joining $u$ to the
constant path at $\left\{ p\right\} =S\cap F_{\varpi }$, up to homotopies, and the choices of such $\mathbf{u}$'s for a given $u$ form a torsor over $%
H_{1}\left( F_{\varpi },\mathbb{Z}\right) \simeq H^{3}\left( Y,\mathbb{Z}%
\right) $. 
\begin{equation*}
f\left( u\right) =\int_{\left[ 0,1\right] ^{2}}\mathbf{u}^{\ast }\omega _{%
\mathcal{J}_{sympl}\left( Y\right) }^{\mathbb{C}}
\end{equation*}%
where $\omega _{\mathcal{J}_{sympl}\left( Y\right) }^{\mathbb{C}}$ is the
canonical holomorphic symplectic form on $\mathcal{J}_{sympl}\left( Y\right) 
$.

Critical points of $f$ are lifts of $p$ in $\widetilde{\Omega }%
_{S\rightarrow F_{\varpi }}\left( \mathcal{J}_{sympl}\left( Y\right) \right) 
$, thus $Crit\left( f\right) $ is a $H^{3}\left( Y,\mathbb{Z}\right) $%
-torsor. Beasseau has a surprising conjecture \cite{Bousseau} which implies
that the category $_{A}Hom_{X}\left( S,F_{\varpi }\right) $ contains $%
DT_{\left( Y,\varpi \right) ,\Omega }^{SLag}\left( \gamma \right) $'s for
ALL $\gamma \in H_{3}\left( Y,\mathbb{Z}\right) \simeq H^{3}\left( Y,\mathbb{%
Z}\right) $. Concretely, objects in $_{A}Hom_{X}\left( S,F_{\varpi }\right) $
are critical points of $f$. Given two such critical point $p_{1}$ and $p_{2}$%
, denote $\gamma =p_{2}-p_{1}\in H_{3}\left( Y,\mathbb{Z}\right) $, the
conjecture asserts that the Hom space between $p_{1}$ and $p_{2}$ in $%
_{A}Hom_{X}\left( S,F_{\varpi }\right) $ is given be $H^{\mathrm{DT, SLag}}_{\left( Y,\varpi\right) ,\Omega }\left( \gamma \right) $. It would be interesting to
related Bousseau's conjecture with the DT-branes that we discussed earlier.

\bigskip


\bigskip

\begin{thebibliography}{99}
\bibitem{Abouzaid} M. Abouzaid. \textit{The family Floer functor is faithful}%
. J. Eur. Math. Sco. 19 (2017), no. 7, 2139-2217.

\bibitem{Bousseau} P. Bousseau. \textit{Holomorphic Floer theory and
Donaldson-Thomas invariants, }Proc. Sympos. Pure Math., 107 AMS, Providence,
RI, 2024, 45--78.

\bibitem{BLPW} T. Braden, A. Licata, N. Proudfoot, and B. Webster. \textit{%
Quantizations of conical symplectic resolutions II: category O and
symplectic duality}. Asterisque 384 (2016). with an appendix by I. Losev,
pp. 75--179.

\bibitem{BFN} A. Braverman, M. Finkelberg, and H. Nakajima. \textit{Towards
a mathematical definition of Coulomb branches of 3-dimensional N = 4 gauge
theories, II}. Adv. Theor. Math. Phys. 22.5 (2018), pp. 1071--1147.

\bibitem{Bridgeland} T. Bridgeland. \textit{Stability conditions on triangulated categories}. Annals of Math. (2007) 166 (2), 317-345.

\bibitem{CdGP} P. Candelas, X. de la Ossa, P. Green and L. Parkes. \textit{A
pair of Calabi--Yau manifolds as an exactly soluble superconformal field
theory}. Nuclear Physics B. (1991) 359 (1): 21--74.

\bibitem{Chan Leung} K. F. Chan and N. C. Leung. \textit{SYZ mirrors in
non-Abelian 3d mirror symmetry}. (2024) arXiv: 2408.09479.

\bibitem{Chan Leung 3d} K. F. Chan and N. C. Leung. \textit{3d mirror
symmetry is mirror symmetry}. (2024). arXiv: 2410.03611.

\bibitem{Chan Leung SYZ} K. F. Chan and N. C. Leung, in preparation.

\bibitem{CLL} K.W. Chan, S.-C. Lau and N.-C. Leung. \textit{SYZ mirror
symmetry for toric Calabi-Yau manifolds}. Journal of Differential Geometry
90 (2012), no. 2, 177--250.

\bibitem{Cho Hong Lau} C.H. Cho, H. Hong and S.C. Lau. \textit{Gluing
localized mirror functors}, J. Diff. Geom. 126 (2024), no. 3, 1001-1095.

\bibitem{DT} S.K. Donaldson, R.P. Thomas. \textit{Gauge theory in higher dimensions.} The geometric universe (Oxford, 1996), Oxford University Press, pp. 31–47.

\bibitem{Donovan Wemyss} W. Donovan and M Wemyss. \textit{Stringy K\"{a}hler
moduli, mutation and monodromy. }Journal of Differential Geometry (2025) 129
(1), 115-164.

\bibitem{Douglas} M. Douglas, B. Fiol, C. R\"{o}melsberger. \textit{%
Stability and BPS branes}, hep-th/0002037.

\bibitem{Freed} D. Freed. \textit{Special Kahler manifolds}, CMP. 203
(1999), no. 1, 31-52.

\bibitem{FOOO book} K. Fukaya, Y.-G. Oh, H. Ohta, K. Ono. \textit{Lagrangian
intersection Floer theory: anomaly and obstruction}. Part I.and II AMS/IP
Studies in Advanced Mathematics, 46.1 and 46.2. AMS

\bibitem{GMW} D. Gaiotto, G. Moore and E. Witten. \textit{An introduction to
the web-based formalism, }Confluentes Math. 9 (2017), no. 2, 5--48.

\bibitem{Gammage Hilburn} B. Gammage and J Hilburn. \textit{Hypertoric
2-Categories and Symplectic Duality, }Communications in Mathematical Physics
(2026) 407 (3), 55.

\bibitem{Givental} A. Givental, Alexander. \textit{A Mirror Theorem for
Toric Complete Intersections}. Topological Field Theory, Primitive Forms and
Related Topics. (1998) pp. 141--175. 

\bibitem{Greene Plesser} B. Greene and R. Plesser. \textit{Duality in
Calabi--Yau moduli space}. Nuclear Physics B. (1990) 338 (1): 15--37.

\bibitem{Intriligator Seiberg} K. Intriligator and N. Seiberg. \textit{%
	Mirror symmetry in three-dimensional gauge theories}. Physics Letters B.
(1996) 387 (3): 513--519

\bibitem{Joyce} D. Joyce. \textit{Conjectures on Bridgeland stability for
Fukaya categories of Calabi-Yau manifolds, special Lagrangians, and
Lagrangian mean curvature flow, }EMS Surv. Math. Sci. 2 (2015), no. 1, 1--62.

\bibitem{KKS} M. Kapranov, M. Kontsevich and Y. Soibelman. \textit{Algebra
	of the infrared and secondary polytopes. }Adv. Math. 300 (2016), 616--671.

\bibitem{Kapranov} M. Kapranov and V. Schechtman. \textit{Perverse schobers}%
, arXiv:1411.2772.

\bibitem{KRS} A. Kapustin, L. Rozansky, and N. Saulina. \textit{%
Three-dimensional topological field theory and symplectic algebraic
geometry. I.} Nuclear Phys. B 816.3 (2009), pp. 295--355.

\bibitem{Kapustin Vyas} A. Kapustin and K. Vyas. \textit{A-models in three
and four dimensions}, preprint, hep-th/1002.4241.

\bibitem{Kontsevich ICM} M. Kontsevich. \textit{Homological algebra of
mirror symmetry.} Proceedings of the International Congress of
Mathematicians, Vol. 1, 2 (Zurich, 1994), 120--139, Birkhauser, Basel, 1995.

\bibitem{KS} M. Kontsevich and Y. Soibelman. \textit{Wall-crossing
structures in Donaldson-Thomas invariants, integrable systems and mirror
symmetry, }Lect. Notes Unione Mat. Ital., 15Springer, Cham, 2014, 197--308.

\bibitem{LYZ} N.C. Leung, S.-T. Yau and E. Zaslow. \textit{From special
Lagrangian to Hermitian-Yang-Mills via Fourier-Mukai transform}, ATMP 4
(2000), no. 6, 1319-1341.

\bibitem{LLY} B. Lian, K.F. Liu and S.-T. Yau. \textit{Mirror principle, I}.
Asian Journal of Mathematics. (1997) 1 (4): 729--763.

\bibitem{PPTV} T. Pantev, B. Toen, M. Vaquie and G. Vezzosi. \textit{Shifted
symplectic structures}, Publ. Math. Inst. Hautes Etudes Sci. 117 (2013),
271-328.

\bibitem{Pascaleff} J. Pasceleff and N. Sibilla. \textit{Speculations on
higher Fukaya categories}, preprint, arXiv:2503.20906.

\bibitem{Rozansky Witten} L. Rozansky and E. Witten. \textit{Hyperkahler
geometry and invariants of three manifolds}, Selecta Math. 3 (1997), no.3,
401-458.

\bibitem{Seiberg Witten} N. Seiberg and E. Witten. \textit{Gauge dynamics
and compactification to three dimensions}, The mathematical beauty of
physics (Saclay, 1996), Adv. Ser. Math. Phys., vol. 24, World Sci. Publ.,
River Edge, NJ, 1997, pp. 333--366.

\bibitem{Sheridan} N. Sheridan. \textit{Homological Mirror Symmetry for
Calabi-Yau hypersurfaces in projective space,} Inventiones Mathematicae, 199
(2015), no. 1, 1--186.

\bibitem{Spenko Van Den Bergh} \v{S} \v{S}penko, M Van den Bergh. \textit{A
class of perverse schobers in Geometric Invariant Theory}: Selecta
Mathematica 32 (2), 19

\bibitem{SYZ} A. Strominger, S.-T. Yau and E. Zaslow. \textit{Mirror
symmetry is T-duality}. Nuclear Phys. B 479 (1996), no. 1-2, 243--259.

\bibitem{Thomas Yau} R. Thomas and S.-T. Yau. \textit{Special Lagrangians,
stable bundles and mean curvature flow}. Communications in Analysis and
Geometry. (2002) 10 (5): 1075--1113.

\bibitem{Tian} G\textit{. }Tian. \textit{ Smoothness of the universal
deformation space of compact Calabi-Yau manifolds and its Petersson-Weil
metric}, Mathematical aspects of string theory, 1987, Adv. Ser. Math. Phys.,
vol. 1, pp. 629--646 

\bibitem{Todorov} A Todorov. \textit{The Weil-Petersson geometry of the
moduli space of SU(n\TEXTsymbol{>}=3) Calabi Yau manifolds. I}, Comm. Math.
Phys., 1989, vol. 126, pp. 325--346
\end{thebibliography}
\end{document}